\documentclass[twocolumn]{aastex631}

\usepackage{amsmath}

\begin{document}

\title{Disentangling AGN Feedback and Sloshing in the Perseus Cluster with \textit{XRISM}: Insights from Simulations}

\author[0000-0001-6411-3686]{Elena Bellomi}
\affiliation{Center for Astrophysics $\vert$ Harvard \& Smithsonian, 60 Garden St., Cambridge, MA 02138, USA}

\author[0000-0003-3175-2347]{John A. ZuHone}
\affiliation{Center for Astrophysics $\vert$ Harvard \& Smithsonian, 60 Garden St., Cambridge, MA 02138, USA}

\author{Nhut Truong}
\affiliation{Center for Space Sciences and Technology, University of Maryland, Baltimore County (UMBC), Baltimore, MD, 21250 USA}
\affiliation{NASA / Goddard Space Flight Center, Greenbelt, MD 20771, USA}
\affiliation{Center for Research and Exploration in Space Science and Technology, NASA / GSFC (CRESST II), Greenbelt, MD 20771, USA}

\author[0000-0001-7630-8085]{Irina Zhuravleva}
\affiliation{Department of Astronomy \& Astrophysics, University of Chicago, 5640 S Ellis Ave., Chicago, IL, 60637, USA}

\author{Rainer Weinberger}
\affiliation{Leibniz-Institut f\"{u}r Astrophysik Potsdam (AIP), An der Sternwarte 16, D-14482 Potsdam, Germany}

\author[0000-0002-7275-3998]{Christoph Pfrommer}
\affiliation{Leibniz-Institut f\"{u}r Astrophysik Potsdam (AIP), An der Sternwarte 16, D-14482 Potsdam, Germany}

\author[0000-0001-5888-7052]{Congyao Zhang}
\affiliation{Department of Theoretical Physics and Astrophysics, Masaryk University, Brno 61137, Czechia}
\affiliation{Department of Astronomy \& Astrophysics, University of Chicago, 5640 S Ellis Ave., Chicago, IL, 60637, USA}

\author[0000-0002-7726-4202]{Annie Heinrich}
\affiliation{Department of Astronomy \& Astrophysics, University of Chicago, 5640 S Ellis Ave., Chicago, IL, 60637, USA}

\author[0009-0002-2669-9908]{Mateusz Ruszkowski}
\affiliation{Department of Astronomy, University of Michigan, Ann Arbor, MI 48109, USA}

\author{Brian McNamara}
\affiliation{Department of Physics \& Astronomy, Waterloo Centre for Astrophysics, University of Waterloo, Ontario N2L 3G1, Canada}

\author[0000-0001-7271-7340]{Julie Hlavacek-Larrondo}
\affiliation{Département de Physique, Université de Montréal, Succ. Centre-Ville, Montréal, Québec, H3C 3J7, Canada}
\affiliation{Centre de recherche en astrophysique du Québec (CRAQ)}

\author{M. Gendron-Marsolais}
\affiliation{Département de physique, de génie physique et d’optique, Université Laval, Québec (QC), G1V 0A6, Canada}

\author[0000-0002-2478-5119]{Benjamin Vigneron}
\affiliation{Département de Physique, Université de Montréal, Succ. Centre-Ville, Montréal, Québec, H3C 3J7, Canada}
\affiliation{Centre de recherche en astrophysique du Québec (CRAQ)}

\begin{abstract}
High-resolution X-ray spectroscopy with XRISM has revealed complex, non-monotonic velocity dispersion profiles in the Perseus cluster, pointing to a complex interplay between at least two physical drivers of motions caused by dynamical processes within the intracluster medium (ICM). To further explore this conclusion, 
we perform a suite of idealized, controlled simulations targeting the relative roles of merger-induced sloshing and active galactic nucleus (AGN) feedback. Our models systematically isolate and combine these mechanisms to predict observable velocity profiles and X-ray line shapes, providing direct comparison to XRISM and Hitomi data. We find that neither sloshing nor AGN activity alone can reproduce the observed velocity dispersion profile; only their combined action matches the elevated dispersions both at the cluster core and outskirts. Power-spectrum analysis reveals distinct spatial signatures: sloshing generates large-scale coherent motions, while AGN feedback injects turbulence and broadens the velocity spectrum at small scales, especially in the core. By forward-modeling spectral line profiles, we show how these dynamics imprint unique observational signatures on X-ray emission. Our results underscore the necessity of accounting for both large-scale and small-scale drivers of gas motions in the ICM when interpreting high-resolution spectroscopic data, and provide guidance for the analysis of forthcoming XRISM observations.
\end{abstract}

        
\section{Introduction} \label{sec:intro}
        A significant portion of the universe's baryonic matter is found in the form of hot, X-ray-emitting gas, which permeates massive galaxies, groups, and galaxy clusters, and is dispersed throughout the interstellar and intergalactic media. Galaxy clusters, as the largest gravitationally bound structures, play a crucial role in our understanding of cosmological models. They form hierarchically through a series of mergers---events that release energy at scales second only to the Big Bang, reaching up to $10^{64} - 10^{65}$ erg \citep{sarazin_2002}. These mergers imprint themselves onto the intracluster medium (ICM) through distinct X-ray surface brightness discontinuities, manifesting as shocks and cold fronts \citet{markevitch_shocks_2007,zuhone_cold_2016,zuhone_merger_2022}.
        Cold fronts are contact discontinuities in X-ray surface brightness, separating a region of denser and cooler gas from hotter, less dense plasma. These phenomena often result from sloshing motions within the cluster's gravitational potential well, predominantly shaped by dark matter. In ``cool-core'' clusters, where the core is dense and low-temperature, gas sloshing is typically initiated by the gravitational passage of a subcluster. This motion displaces the low-entropy, cold core gas from the potential minimum, initiating oscillatory motions that develop into distinct spiral-shaped cold fronts. As these spirals move outward from the cluster center, they can form long-lived structures that owe their shape to the angular momentum from off-center encounters.
        
        The Perseus Cluster, as the brightest galaxy cluster in the X-ray sky, has been the focus of numerous detailed observations, enabling the discovery of its rich and complex array of features. It exhibits spiral sloshing cold fronts at small radii, along with AGN-inflated X-ray cavities (or ``bubbles'') \citep{boehringer_1993}, ripples \citep{fabian_2003,fabian_2006}, and features such as the ``bay'' \citep{walker_is_2017}, potentially linked to Kelvin-Helmholtz instabilities (hereafter KHI). At larger radii, additional X-ray discontinuities, like ancient cold fronts and/or shocks, begin to appear even near the cluster's virial radius \citep{churazov_xmm-newton_2003, fabian_wide_2011, simionescu_large-scale_2012, walker_is_2017, walker_split_2018, walker_is_2022}.
        
        Understanding the velocity field of the ICM is essential for interpreting the origin and evolution of the observed structures within galaxy clusters, such as the above-mentioned cold fronts, shocks, and AGN-inflated cavities. Measurements of the line-of-sight velocity and turbulent motions provide direct insight into the dynamical state of the ICM, helping disentangle the effects of active galactic nucleus (AGN) feedback, cluster mergers, and sloshing motions. 

        Observationally, determining gas velocities in the dense cores of clusters has historically been challenging due to limitations in both spectral resolution and sensitivity. The advent of high-resolution X-ray spectroscopy, first demonstrated by the Hitomi mission and now advanced by XRISM's Resolve instrument, has markedly improved this situation. Resolve is a micro-calorimeter spectrometer with $\sim$5~eV spectral resolution \citep{Ishisaki_2022}.  This enables the mapping of ICM velocity fields through measurements of Doppler shifts and line broadening in the emission lines of heavy ions such as iron. As a result, it is now possible to directly quantify bulk motions and turbulence in the ICM across a range of spatial scales, opening a new window for testing theoretical models of cluster dynamics.

        Recent XRISM observations of the Perseus Cluster \citep{xrism_perseus_2025} have produced the first high-resolution, spatially resolved kinematic radial mapping of the cluster core  (see also \citealt{zhang_perseus_2025}). These data reveal a non-monotonic, ``W-shaped'' radial profile in the velocity dispersion, with elevated gas motions both in the central $\sim$60~kpc and at larger radii. This complex structure suggests the presence of at least two distinct processes driving gas motions on different spatial scales: one acting on small scales and likely associated with supermassive black hole feedback in the core, and another due to large-scale merger activity affecting the cluster outskirts.

        Despite the advancements in observational capabilities, interpreting the complex gas motions within galaxy clusters remains a significant challenge. A fundamental question persists: are the observed velocities within clusters primarily driven by active galactic nucleus (AGN) feedback mechanisms, or are they largely a consequence of merger-induced dynamics? AGN feedback involves the interaction between energy output from the central supermassive black hole and the surrounding intracluster medium (ICM), often resulting in powerful jets and outflows that can drive gas motions. Previous work has demonstrated the velocity signatures one would expect to see from AGN feedback in terms of both line shifts and line broadening \citep[e.g.][]{Bruggen2005,Shang2012,Lau2017,Bourne_2017,Ehlert_2021}. On the other hand, mergers between galaxy clusters disturb the ICM through gravitational interactions, generating turbulence and large-scale flows that also manifest in emission line signatures \citep[e.g.][]{zuhone_astroh_2016,zuhone_hitomi_2018,Biffi2022}.

        This dichotomy in potential driving mechanisms complicates the interpretation of observed velocity fields. Differentiating between AGN-driven and merger-driven motions is crucial because it influences our understanding of energy transfer processes, cluster evolution, and the role of AGN feedback in regulating the thermal state of cluster cores.

        To physically interpret the complex velocity structures and line profiles revealed by XRISM in Perseus, we conduct a series of idealized numerical simulations designed to separate the roles of merger-induced sloshing and AGN feedback. By modeling both processes $-$ independently and in combination $-$ under Perseus-like conditions, we can predict their distinct kinematic signatures across relevant spatial scales and compare these to the new XRISM observations. This approach enables us to assess which physical mechanisms are necessary to reproduce the observed velocity dispersion profiles and spectral line shapes, thus providing a theoretical framework for understanding the multi-scale drivers of gas motions in the cluster core. 

        We note that a complementary study (Truong et al., in prep.) analyzes the XRISM velocity dispersion profile using cosmological hydrodynamical simulations, where cluster gas motions arise naturally from mergers, accretion, and AGN feedback. Combined with the idealized experiments presented here, these results provide a unified theoretical framework for interpreting the origins of turbulence and bulk motions in the Perseus ICM.

        This paper is organized as follows. In Section~\ref{sec:method}, we outline the assumed physics and numerical setup of our simulations. Section~\ref{sec:results} presents the results, including detailed comparisons between the different physical scenarios explored. In Section~\ref{sec:conclusions}, we summarize our findings and present our conclusions. Throughout, we assume a flat $\Lambda$CDM cosmology with $h$ = 0.71, $\Omega_m$ = 0.27, and $\Omega_\Lambda$ = 0.73.

\section{Methods: simulations and tools}\label{sec:method}

    \subsection{Initial Conditions}

    We model the merger of a galaxy cluster and an infalling subcluster, each composed of fully ionized gas in hydrostatic equilibrium within a dominant dark matter (DM) halo. The ICM of both clusters are characterized by an adiabatic index of $\gamma = 5/3$ and a mean molecular weight of $\mu = 0.6$. The primary cluster is initialized with a total mass of $M = 5.9 \times 10^{14} M_\odot$, following a super-NFW (sNFW) density profile for the dark matter halo, defined as

    \begin{equation}
    \rho_{\rm sNFW}(r) = \frac{3M}{16\pi a^3} \frac{1}{x(1 + x)^{5/2}}, \quad x = \frac{r}{a},
    \end{equation}
    where $a = 389.5$~kpc represents the scale radius. The gas density of the main cluster is modeled as the sum of a classic $\beta$-model and a modified $\beta$-model profile from \citet{vikhlinin_chandra_2006}, representing the gas distribution in both the core and outer regions. The electron density profile is given by

    \begin{equation}
    \begin{split}
    n_e(r) &= \frac{n_{e,c1}}{[1 + (r / r_{c1})^{2}]^{1.5\beta_1}}\\
    &+ \frac{n_{e,c2}}{[1 + (r / r_{c2})^2]^{1.5 \beta_2} [1 + (r / r_s)^\gamma]^{\epsilon / 2\gamma}},
    \end{split}
    \end{equation}
    where $n_{e,c1} = 4.5 \times 10^{-2}$ cm$^{-3}$ and $n_{e,c2} = 4 \times 10^{-3}$ cm$^{-3}$ are the central electron densities of the two components. The core radii are $r_{c1} = 55$ kpc and $r_{c2} = 180$ kpc, while the outer profile is characterized by $r_s = 1800$ kpc, $\gamma = 3$, and $\epsilon = 3$. The slopes of the inner profiles are set by $\beta_1 = 1.2$ and $\beta_2 = 0.6$, resulting in a gas density profile that closely matches that observed in Perseus \citep{zhuravleva_resonant_2013,urban_azimuthally_2014}.

    We initialize the gas with a turbulent, isotropic magnetic field following the procedure of \citet{zuhone_how_2020}. A Kolmogorov-spectrum magnetic field is generated on a uniform grid and rescaled so that the average plasma beta, $\beta = p_{\rm th}/p_B$, is approximately constant across the domain. We adopt here $\beta = 200$. 
    
    \subsection{Merger Simulation}
    
    The evolution of the merging system is modeled using the moving-mesh magnetohydrodynamics (MHD) code AREPO \cite{springel_e_2010,pakmor_2016,weinenberger_arepo_2020}, which solves the equations of ideal MHD while allowing for adaptive spatial resolution. This approach accurately captures the complex gas dynamics and the evolution of magnetic fields during the merger process. Both dark matter and gas components are evolved, with the dark matter treated as a collisionless fluid under a gravitational potential and the gas governed by the MHD equations, which are evolved using the Powell divergence-cleaning scheme \citep{powell_solution_1999}, as implemented in previous moving-mesh studies \citep{pakmor_simulations_2013, marinacci_first_2018}, which controls numerical $\nabla \cdot \mathbf{B}$ errors via additional source terms in the MHD equations.
    The refinement of the gas cells are set up so that the mean gas mass is $1.08 \times 10^{7}$~M$_{\odot}$.
    A given cell is refined or derefined if its mass deviates by more than a factor of 2 above or below this value. 
    
    The merger is set up by assigning a relative velocity to the subcluster, allowing it to fall toward the main cluster. The subcluster is initially located at a distance of 3000~kpc along the $+y$-axis from the center of the main cluster, with a relative velocity of $v_{\rm sub} = 1.1 V_{200c} \approx 1380$ km~s$^{-1}$, where $V_{200c}$ is the main cluster's circular velocity, although higher velocities up to 2000 km~s$^{-1}$ are also considered. For our fiducial model, we set the mass of the subcluster such that the mass ratio is $R = M_{\rm sub} / M_{\rm main} = 1:5$. The gas density profile of the subcluster follows a \citet{vikhlinin_chandra_2006} profile with $r_s = 300$~kpc, $r_{c2} \sim 1000$~kpc, and $\beta_s = 0.67$. The subcluster's infall angle relative to the line of centers is $\theta \approx 25^{\circ}$, as motivated by the distribution of merger angles observed in cosmological simulations and previous studies of Perseus large-scale cold fronts \citep{bellomi_2024}. This corresponds to an impact parameter of $\sim 1268$~kpc. This configuration produces gravitational interactions that drive cold front formation, producing large-scale cold fronts at the radii seen in Perseus, and generates a spiral structure near the core. The magnetic field initial condition described above is cleaned in Fourier space to remove non-divergence-free components and then interpolated onto the simulation mesh. 
    
    \subsection{AGN Jet Feedback Implementation}
    Multiple studies indicate that the central AGN in Perseus (NGC 1275) operates in an episodic mode, with outbursts inflating X-ray cavities in the surrounding intracluster medium (e.g., \citealt{birzan_2004, rafferty_2006, fabian_2006, fabian_2011}). The mechanical power associated with these outbursts, as estimated from the energetics of the observed cavities, ranges from $P_{\rm cav} \sim 1 \times 10^{44}$ to $1 \times 10^{45}$~erg~s$^{-1}$. Jet power estimates based on cavity enthalpy alone give a value of $\sim 1.5 \times 10^{44}$~erg~s$^{-1}$ \citep{rafferty_2006}. However, this estimate does not include additional energy released in the post-shock gas and in sound waves, nor does it include particle energy that may have leaked away from the bubbles. Accounting for these extra factors, \citet{fabian_2011} estimate the average jet power over the past $\sim$100~Myr could be as large as $\sim10^{45}$~erg~s$^{-1}$. As we will show, our simulations indicate that jet power on the higher end of this estimate would be required to produce the central velocity dispersion observed with XRISM.
    
    Analyses of bubble ages and spacing suggest a typical on/off duty cycle of a few tens of Myr between major outbursts. In our simulations, we adopt a jet power $P_{\rm jet}$ of either $3 \times 10^{44}$ or $1 \times 10^{45}$~erg~s$^{-1}$, and implement episodic activity with 100~Myr of jet activity followed by a 30~Myr quiescent phase, consistent with these observational constraints and also with simulations of self-regulated AGN jet feedback in a Perseus-like cluster \citep{ehlert_self-regulated_2023}. This setup allows the formation of round, buoyant bubbles while preventing excessive cooling in the core. While a detailed exploration of the duty cycle's impact is beyond the scope of this study, our approach reflects the observed intermittent nature of AGN feedback in Perseus and similar clusters. 
    
    We model active galactic nucleus (AGN) feedback using the black hole and jet models described in \citet{weinberger_simulating_2017,weinberger_2023}. Black holes inject energy through kinetically dominated, low-density collimated, bipolar outflows in pressure equilibrium with their surroundings.
    In our simulations, the jets are injected bidirectionally along the $y$-axis, with kinetic, thermal, and magnetic energy deposited into two small spherical regions a few kpc away from the black hole particle's position. This jet orientation results in their residing in the same plane as the sloshing motions from the merger. In Perseus, multiple pairs of cavities are situated on opposite sides of the cluster center, and the cold fronts appear in large, azimuthally-connected spirals \citep{fabian_2003,fabian_2006,fabian_2011}. Both of these features are indicators that these two processes share a nearly common plane that has a significant component within the plane of the sky. The fact that the buoyant cavities do not appear on a single line on either side of the cluster center may indicate that the sloshing motions have pushed them away from the jet axis, where they were launched \citep[][]{Fabian_2022,paola_2024}.  
    
    The black hole particle, which serves as the jet injection site, is placed at the cluster potential minimum and has a mass of $M_\mathrm{BH} = 1 \times 10^9$ M$_\odot$. This mass is large enough to prevent the particle from being significantly displaced by interactions with dark matter particles, ensuring it remains close to the cluster potential minimum throughout the simulation.
    The jet material is injected in a sphere of 1 kpc around the central black hole and has a density of $6 \times 10^{-7}$ in code units (corresponding to $\sim 4 \times 10^{-28}$ g/cm$^{3}$ in physical units). The injected jets are approximately two orders of magnitude lighter than the surrounding gas.
    A passive tracer field is injected by the jets, and if the mass fraction associated with this tracer exceeds $2 \times 10^{-3}$, the cell is refined.
    
    The energy injected by the jet is predominantly kinetic, with magnetic and thermal pressures set to be equal within the jet region, corresponding to a magnetic-to-thermal energy ratio of $\beta_\mathrm{jet} = 1$. The injected magnetic field is purely toroidal. Jet material is traced using a passive scalar field that is advected with the fluid throughout the simulation. This scalar allows us to distinguish jet-entrained material from the ambient ICM. 
    
    The lobes generated by the AGN jets also contain cosmic rays (CRs), which are included in the treatment of the gas dynamics using the two-fluid approximation \citep{Pakmor2016,Pfrommer2017}. Similar to the magnetic field, the ratio of CR energy to thermal energy is unity. CRs have been shown to provide a stable balance where Alfv\'en wave heating offsets radiative cooling \citep{jacob_2017_I,jacob_2017_II,ehlert_simulations_2018}. \citet{Ruszkowski2017} showed that CR streaming, diffusion, and Coulombic and hadronic interactions provide important sources of heating. For simplicity, for the purposes of this work, and since our primary focus is on the velocity structure and the effects of AGN-driven bubbles on velocity dispersion, we do not model additional processes affecting cosmic ray evolution, such as diffusion, streaming, (re)acceleration, or energy losses.

    Our simulations explicitly track the spatial distribution and evolution of metallicity in the ICM. For the initial metallicity profile, we adopt the parametrization of \citet{mernier_2017}, motivated by X-ray observations of cool-core clusters such as Perseus. In all of our simulations, the metallicity is treated as a passive scalar: its evolution is governed solely by hydrodynamical advection and mixing, capturing the redistribution of metals due to bulk gas motions. We employ the Galaxy Formation Model (GFM) framework in AREPO \citep{Vogelsberger_2013} to include metal-dependent radiative cooling (which is active throughout the simulation), with rates computed based on pre-tabulated tables accounting for metal-line emission across a range of temperatures and abundances. While star formation is formally enabled within the GFM to prevent excessive gas buildup, both stellar evolution and ongoing chemical enrichment processes are disabled in these runs. This choice allows us to isolate the hydrodynamics of the ICM and the interaction of AGN-driven outflows with the ambient plasma, without introducing additional sources of metals during the course of the simulations.

    Although we adopt the AGN feedback model of \citet{weinberger_simulating_2017,weinberger_2023}, this choice is not unique. Several plausible AGN feedback prescriptions exist, and our aim here is not to favor a specific implementation but to explore how AGN-driven outflows, in contrast to sloshing-driven motions, influence the resulting gas kinematics. We use this model because its kinetic mode provides a well-tested, directionally collimated injection of energy and momentum that captures jet-like behaviour in a consistent subgrid framework, making it particularly suitable for idealized simulations where the impact of AGN-driven flows needs to be isolated and systematically studied. Alternative feedback models could also be used, but this particular model offers a robust and widely adopted choice for this type of experiment.

    \subsection{Simulated X-ray Observables and Diagnostics}\label{sec:linee_profile}

    To enable quantitative comparisons between our simulations and \textit{XRISM}/Resolve observations of the Perseus Cluster, we generate synthetic X-ray observables -- surface brightness maps, velocity field diagnostics, and emission line profiles -- directly from simulation outputs.

    We first compute the X-ray emissivity in the $6$--$8~\mathrm{keV}$ band (surrounding the Fe-K lines used in XRISM observations to measure gas velocities) for each simulation cell using the Astrophysical Plasma Emission Code (APEC) model \citep{smith_collisional_2001}, incorporating the metallicity field evolved in the simulation. 

    For image and map production, we project the emissivity of each gas cell onto the plane of the sky perpendicular to a chosen line of sight (LOS) using a standard SPH cubic spline kernel. The smoothing length $h_i$ for cell $i$ is computed from its Voronoi cell volume $V_i$ by 
    \begin{equation}
    h_i = 2 \left( \frac{3V_i}{4\pi}\right)^{1/3},
    \end{equation}
    ensuring adaptive spatial resolution and conservation of emission \citep{springel_e_2010}.\\

    To probe gas dynamics, we produce maps of the emission-weighted LOS velocity fields by weighting each gas cell's velocity with its X-ray emissivity, as above. The emission-weighted LOS velocity dispersion is calculated as
    \begin{equation}
    \sigma_{\rm los} = \sqrt{ 
          \frac{ \sum_i w_i v_{{\rm los},i}^2}{ \sum_i w_i}
          - 
          \left(\frac{\sum_i w_i v_{{\rm los},i}}{\sum_i w_i}\right)^2
        },
    \end{equation}
    where $v_{\rm los, i}$ is the LOS velocity of the $i$-th cell, and $w_i$ is its emissivity in the band of interest (proportional to the density squared $\rho^2$). These velocity statistics can be directly compared to the moments derived from spectral line profiles observed with the \textit{XRISM}/Resolve microcalorimeter. 

    To synthesize spatially-resolved X-ray emission line profiles, we focus on regions corresponding to the \textit{XRISM}/Resolve field of view -- a square of $60~\mathrm{kpc} \times 60~\mathrm{kpc}$ in the plane of the sky at the Perseus distance. We employ the \texttt{pyXSIM} toolkit\footnote{\url{https://hea-www.cfa.harvard.edu/~jzuhone/pyxsim/}} to generate rest-frame and Doppler-shifted and broadened spectra for gas within this volume, where for the latter case the velocity along the sight line is used.

    For the analysis of velocity structure, we focus on the $6.68$--$6.705~\mathrm{keV}$ energy range, covering the Fe~{\sc xxv} K$\alpha$ line complex, which is a primary tracer of ICM kinematics in clusters such as Perseus \citep{hitomi_2016}.

    We do not convolve the synthetic spectra with the XRISM/Resolve instrumental response. Our comparative analysis focuses on physical trends and relative changes in line profile shapes and widths, rather than exact reproduction of the spectra as processed by the instrument. None of the conclusions reached in this paper require a simulation of the instrumental response.

\section{Results}\label{sec:results}
    \subsection{Velocity Profiles}
        \begin{figure*}[ht!]        
        \plotone{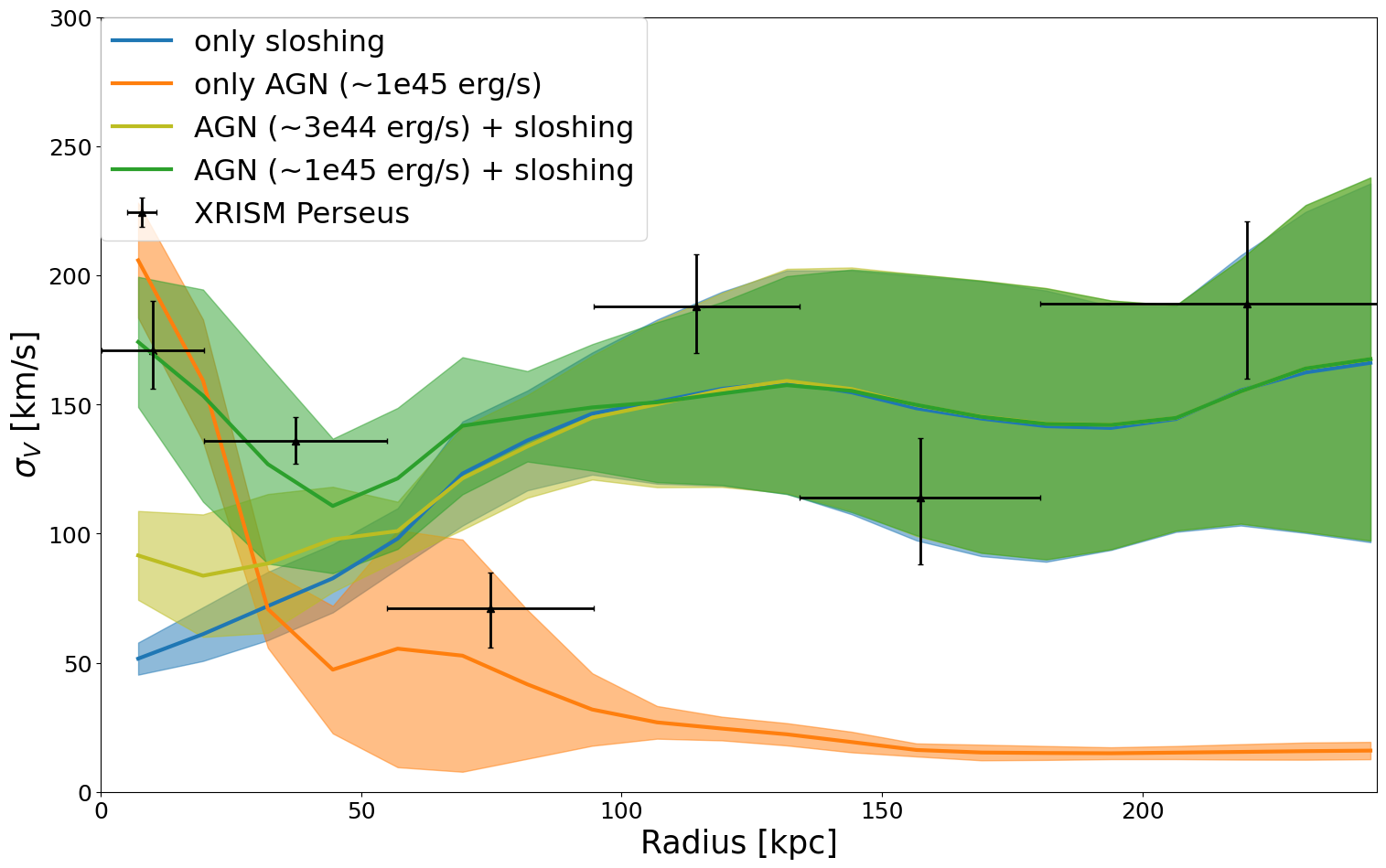}
        \caption{
        The black crosses correspond to the velocity dispersion values and their error bars observed by XRISM \citep{xrism_perseus_2025}. The blue shaded area illustrates results from a sloshing-only simulation, the orange region shows an AGN-only simulation, and the green and light green regions correspond to simulations with AGN jet feedback with different powers included in the same sloshing system. The solid lines correspond to the mean azimuthal values of the emission-weighted velocity dispersion, and the shaded regions indicate the standard deviation from the mean.
        All simulations were initialized identically and are shown at the same epoch, $\sim 1.1$ Gyr after pericenter passage. AGN jet feedback is introduced approximately 1 Gyr after the subcluster pericenter passage and persists for 100 Myr. The merger plane is taken to be the plane of the sky.
        \label{fig:velocity_dispersion}}
        \end{figure*}

        Recent XRISM observations of the Perseus Cluster have revealed a non-monotonic radial profile of the line-of-sight (LOS) velocity dispersion in the ICM (see black crosses in Figure~\ref{fig:velocity_dispersion}). In the inner $\sim$60 kpc, the LOS velocity dispersion rises sharply, reaching values of $\sim$160–200 km~s$^{-1}$ near the central supermassive black hole (SMBH), before dropping to a minimum of $\sim$70 km~s$^{-1}$ at $\sim$70 kpc. Beyond this radius, the velocity dispersion increases again, approaching similar levels to the core out to $\sim$220 kpc. Notably, local minima at $\sim$50 and $\sim$140 kpc coincide with prominent spiral structures in the residual X-ray surface brightness image, strongly suggesting a link between the observed kinematic features and merger-driven sloshing motions. This observed profile was interpreted to be a result of two drivers operating in the Perseus core: a small-scale one in the central $\sim$50 kpc, and a large-scale one in the rest of the gas.
        
        We strengthen these conclusions by comparing the observations with idealized simulations of Perseus-like clusters that isolate the effects of sloshing and AGN feedback. Figure~\ref{fig:velocity_dispersion} presents azimuthally averaged radial profiles of the LOS velocity dispersion from four simulations: one with sloshing only (blue), one with AGN feedback only (orange), and two with AGN jets of differing power injected after pericenter passage of the subcluster, hence on top of sloshing. The solid lines represent the mean profile, and the bands indicate the standard deviation from the mean. The velocity dispersion profile from \citet{xrism_perseus_2025} is also shown with black points. The observed profile was taken along one ``arm'' of four pointings by Resolve; our azimuthal sampling provides an estimate of the cosmic variance expected in this physical situation. All snapshots are taken at the same epoch for consistent comparison, $\sim 1.1$ Gyr after pericenter. AGN jet feedback is introduced approximately 1 Gyr after the subcluster pericenter passage and persists for 100 Myr. 

        \textbf{At large radii} ($r \gtrsim 120$~kpc), the observed velocity dispersion matches well with the predictions of sloshing-only simulations. In this regime, the impact of merger-induced gas motions is dominant, and the velocity dispersion remains elevated at $100$--$200$~km~s$^{-1}$. The dispersion profile is nearly identical in all simulation cases that include sloshing, indicating that the dynamics at these scales are largely insensitive to the addition of AGN jet activity. 

        \begin{figure*}[ht!]        \plotone{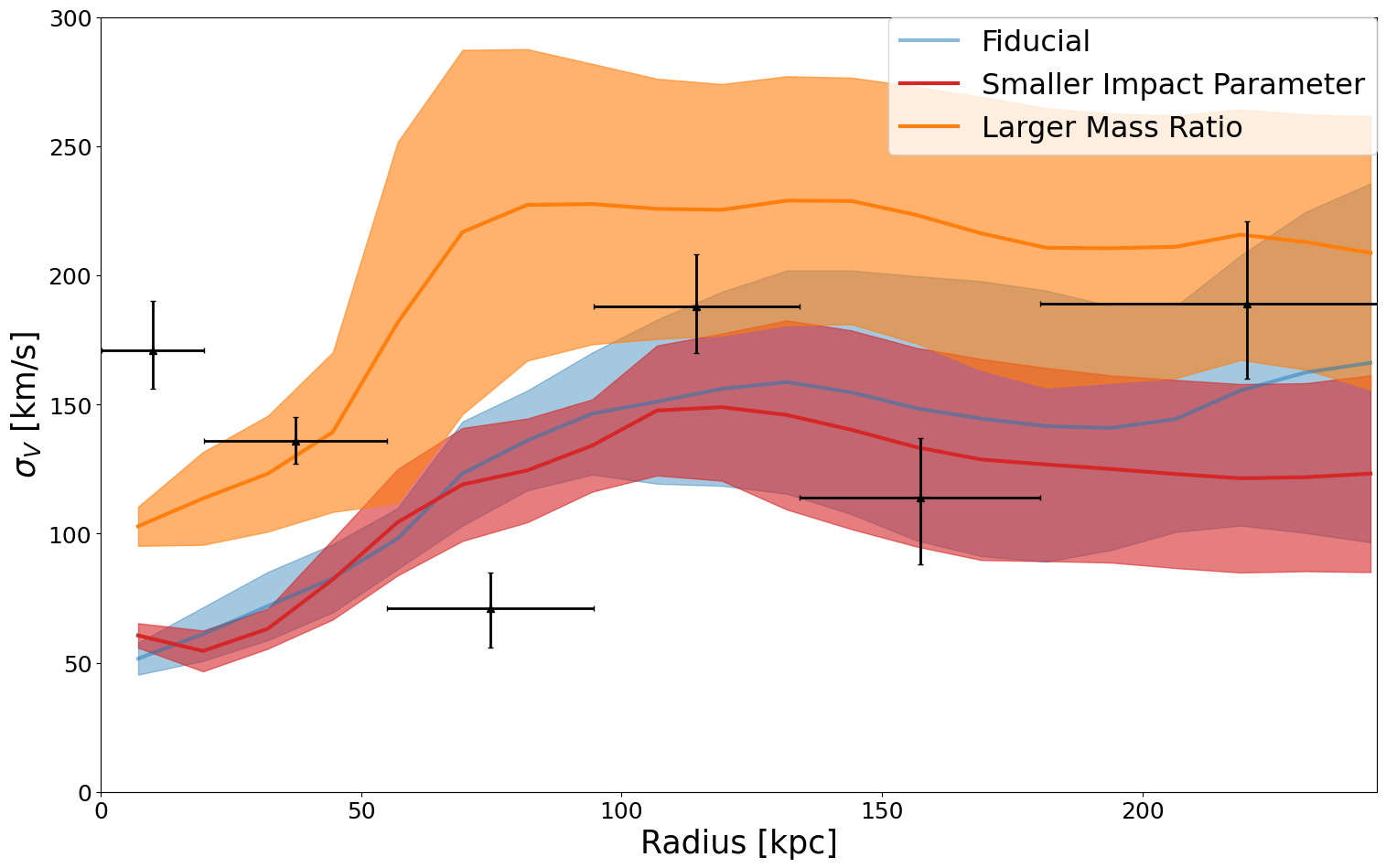}
        \caption{
        Velocity dispersion profiles from simulations exploring different merger configurations compared to the observed values in Perseus \citep[black crosses with error bars; data from][]{xrism_perseus_2025}. The blue shaded region shows results from our fiducial sloshing-only simulation (mass ratio $R = 1:5$, incident angle $\theta = 20^\circ$, which leads to an impact parameter $\sim 1026$ kpc). The orange shaded region illustrates the effect of a more massive subcluster infalling onto the main cluster ($R = 1:2$), which increases the overall velocity dispersion at all radii but still fails to increase the velocity dispersion in the cluster core. The red shaded region corresponds to a simulation with a reduced impact parameter ($\sim 520$ kpc, achieved by setting the incident angle $\theta = 10^\circ$ while keeping the mass ratio fixed).
        The solid lines and shades correspond to the mean values and standard deviation of the emission-weighted velocity dispersion, respectively.
        \label{fig:velocity_dispersion_largerR}}
        \end{figure*}
        
        In the \textbf{central region}, simulations that include AGN feedback uniquely succeed in reproducing the high central velocity dispersions observed by XRISM. 
        When AGN jets are included, the simulations reproduce both the qualitative trend and quantitative amplitude of the observed central velocity dispersion. The jets inflate buoyant cavities that displace the surrounding ICM, generating gas motions that lead to the increase of velocity dispersion within the inner $r \lesssim 50$~kpc. The magnitude of this enhancement scales with the power of the jets: the most energetic case, with jet power $P_{\mathrm{jet}} = 10^{45}$ erg/s, shows the largest increase in dispersion near the center. This indicates that AGN activity plays a dominant role in shaping the gas kinematics in the cluster core. In contrast, simulations with sloshing alone systematically fail to match the core kinematics: here, the central velocity dispersion actually dips, reaching values as low as $\sim50$~km~s$^{-1}$.
        
        This behavior persists even when varying the merger parameters, such as increasing the mass of the infalling subhalo or decreasing the impact parameter to enhance sloshing. While such variations do lead to stronger shear velocities and more evident cold fronts, they still do not predict an increase in the velocity dispersion towards the cluster center, suggesting that sloshing alone cannot generate sufficient gas motions in the core, see Figure~\ref{fig:velocity_dispersion_largerR}. In particular, increasing the subcluster mass (orange shaded region) results in a generally higher velocity dispersion at all radii, as the more massive infalling object imparts a stronger perturbation to the main cluster gas. Nevertheless, the velocity dispersion profile still exhibits a clear dip in the core, failing to match the central values seen in the observations. Reducing the impact parameter by adopting a smaller incident angle for the merger (red region) produces 
        a larger force in the center and leads to a modest increase in the innermost velocity dispersion. Also, the predicted velocity dispersion at larger radii ($r \gtrsim 100$~kpc) is reduced, due to the lower angular momentum for this reduced impact parameter. Otherwise, since the mass ratio in this scenario remains similar to that of the fiducial simulation, the overall velocity dispersion profile remains comparable to the fiducial case. 
        In both cases, the simulated profiles fall significantly short of reproducing the observed central peak. Overall, these tests reinforce the conclusion that sloshing, regardless of merger strength or geometry, cannot on its own explain the high central velocity dispersions measured in Perseus. Thus, another driver of turbulence on a different scale is required to explain the enhanced velocity dispersion in the cluster core, and in Perseus the obvious candidate is the central AGN.

        Turning to the AGN-only simulations, we find that strong AGN activity -- with jet powers up to $10^{45}$~erg~s$^{-1}$ -- can reproduce the high velocity dispersions observed in the cluster core without requiring sloshing from merger events. However, it fails to generate significant motions at larger radii, where the line-of-sight velocity dispersions remain low, typically around $\sim$15~km~s$^{-1}$. This indicates that AGN-driven turbulence primarily stirs the central regions of the cluster, enhancing the velocity dispersion there while leaving the outer ICM largely unaffected over the timescales considered. We also observe that there is a small enhancement in the velocity dispersion between $r \sim 50-100$~kpc, at the edge of the expanding bubble, before decreasing to low dispersion at larger radii. The variance of the velocity dispersion in this radial range is larger, due to the fact that the projected dispersion varies considerably with azimuthal direction (largest along the $y$-axis direction where the jets are fired, and almost zero along the $x$-axis perpendicular to the jets).

        Importantly, the velocity dispersion profile is largely insensitive to the system's inclination with respect to the observer. Unlike LOS bulk velocities -- which can vary significantly with viewing angle -- the velocity dispersion measures the spread in velocities along the line of sight and is therefore robust to moderate changes in projection. This makes it a reliable diagnostic for disentangling internal gas dynamics, and justifies our choice to focus on dispersion in this analysis. This holds even for features such as cold fronts, which are essentially shell-like, expanding structures. Regardless of the viewing angle, the observer still observes a similar shell and samples a similar mix of inbound and outbound velocities along the line of sight, as long as the orientation is not extreme ($\lesssim 50^{\circ}$). This geometric property makes velocity dispersion a reliable diagnostic of internal gas dynamics, justifying our emphasis on this quantity in the present analysis.

        Taken together, these results provide compelling evidence that both merger-driven sloshing and AGN feedback are required -- neither mechanism alone can account for the full velocity dispersion profile in Perseus. Sloshing explains the elevated dispersions at large radii, while AGN feedback is necessary to reproduce the high dispersions observed at the cluster center. 
        Their combined effect can self-consistently account for the overall shape of the observed velocity dispersion profile in the Perseus Cluster, e.g., high velocity dispersions in the core region and at larger radii, but with lower dispersions at intermediate radii.

        It is important to note that the ``W-shaped'' radial profile is not necessarily required to confirm a multi-driver scenario. The two sources of motions may merge at various radii, making the dip in the velocity dispersion less prominent or removing it completely. Even an approximately flat velocity dispersion profile remains consistent with the main conclusion of multiple driving mechanisms, provided that the sloshing driver in Perseus operates on scales of at least $\sim$100 kpc \citet[see Extended Data Fig. 8 in][]{xrism_perseus_2025}.

    \subsection{Velocity Power Spectra}

        \begin{figure*}[ht!]   
        \centering
        \includegraphics[width=1\textwidth]{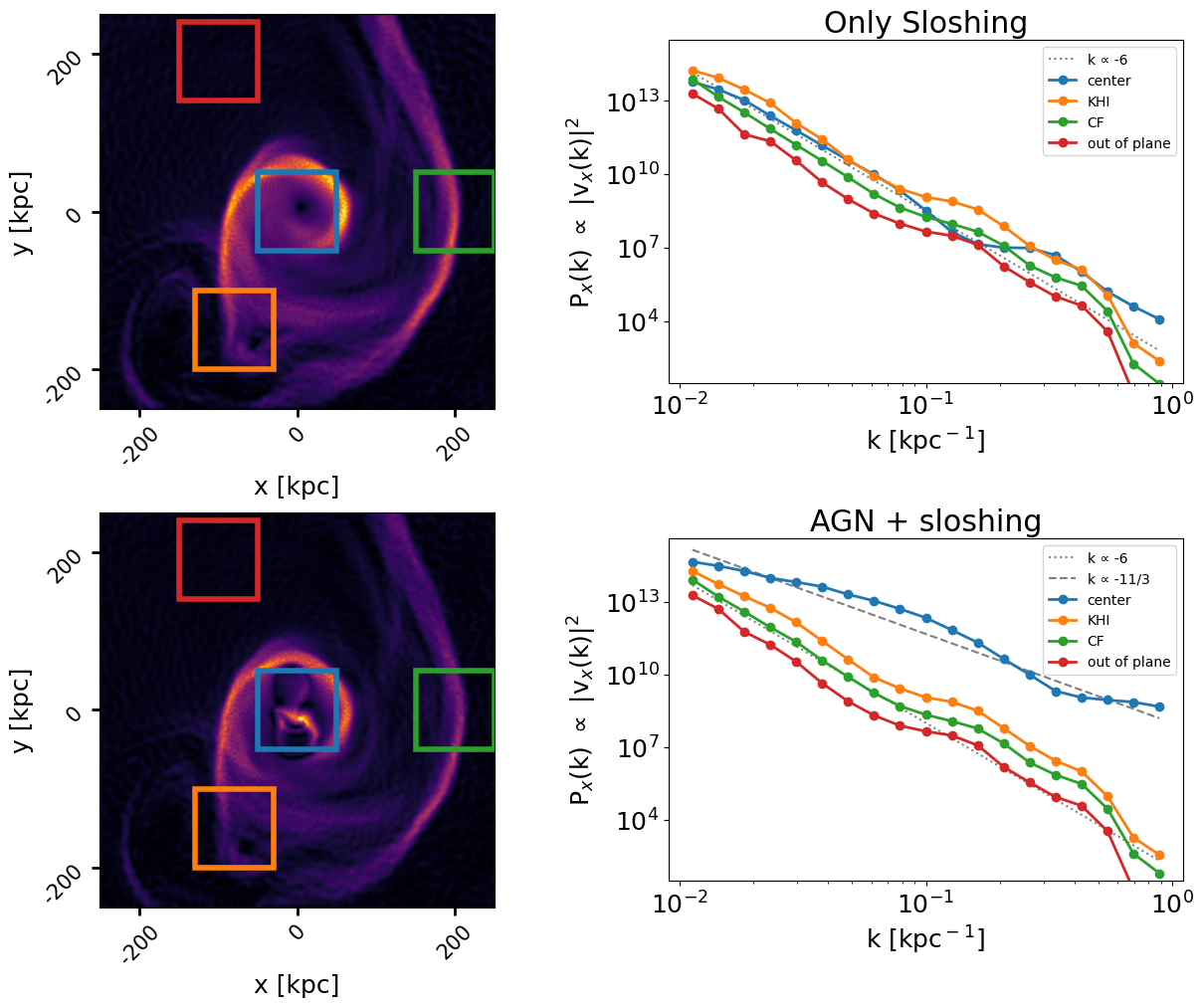}
        \caption{Velocity power spectra in four distinct 100-kpc-wide cubic regions for two simulations. \textit{Left panels}: locations of the cubes overlaid on X-ray surface brightness Gaussian Gradient Magnitude-filtered maps for the sloshing-only simulation (\textit{top}) and the simulation including both sloshing and AGN feedback with jets of power $10^{45}$~erg~s$^{-1}$ (AGN+sloshing, \textit{bottom}). The cubes are labeled in the right panels by location, including the cluster center (blue), a region exhibiting KHI structures (orange), a cold front surface (CF, green), and a region out of the plane (red). \textit{Right panels}: corresponding velocity power spectra for each cube, with colors matching the cube locations. In the sloshing-only case, all regions exhibit similar power spectra dominated by large-scale motions and steep slopes, indicating limited small-scale structure. In contrast, the central region in the AGN simulation shows enhanced power and a flatter slope, consistent with the injection of kinetic energy on smaller scales by AGN-driven activity.
        \label{fig:powerspectra}}
        \end{figure*}

        \begin{figure*}[ht!]   
        \centering
        \includegraphics[width=1\textwidth]{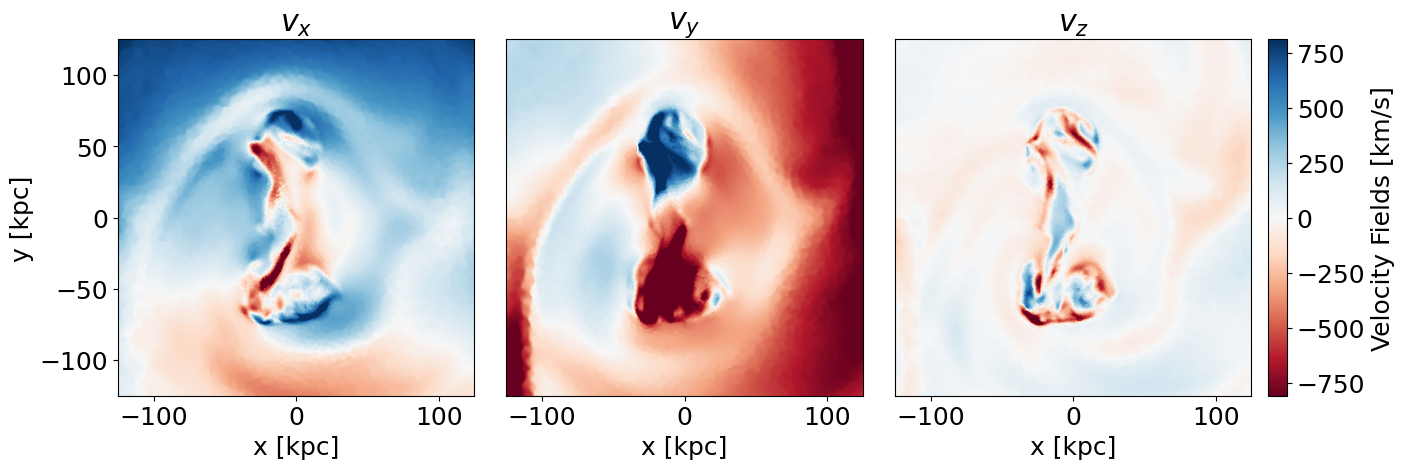}
        \caption{Slices of the $x$, $y$, $z$ components of the velocity field through the cluster core for the simulation including both sloshing and AGN jets with power $1 \times 10^{45}$ erg s$^{-1}$, launched along the $y$-axis.}
        \label{fig:velocity_slices}
        \end{figure*}

        To further characterize the gas motions generated by sloshing and AGN activity, we analyze the velocity power spectra computed within localized cubic regions at various locations in the cluster core and outskirts. This approach enables us to probe how kinetic energy is distributed across spatial scales, and to assess whether different mechanisms preferentially inject power at particular scales or locations.

        Figure~\ref{fig:powerspectra} presents the $x$ component of the 3D power spectrum of the velocity field, $P_k \propto |\tilde{v}_k|^2$, computed in $100~\mathrm{kpc}$-per-side uniformly gridded cubes extracted from four representative regions of the simulated cluster. 
        Prior to computing the spectra, we applied a Tukey filter to each cube to taper the velocity field smoothly to zero at the edges, thereby enforcing periodicity and minimizing aliasing effects. 
        The cubes are chosen to sample areas that exhibit distinct X-ray morphological features in the simulated clusters: (i) the bright central core, (ii) a sharp surface brightness edge associated with a cold front, (iii) a region showing filamentary or vortex-like substructure, indicative of Kelvin-Helmholtz instabilities, and (iv) a comparatively dynamically relaxed region outside the merger plane with no prominent features, as identified in the Gaussian Gradient Magnitude images\footnote{The Gaussian Gradient Magnitude technique highlights gradients in an image by smoothing it with a Gaussian kernel and then taking the magnitude of its gradient \citep{sanders_2016_1,sanders_2016_2,walker_2016}. For our purposes, this highlights features like cold fronts and shocks.} (left panels in Figure~\ref{fig:powerspectra}).
        
        In the sloshing-only simulation, we find that the velocity power spectra obtained in the different cubes are strikingly similar in spectral shape, despite the regions probing disparate physical environments with varying densities and temperatures. This empirical finding demonstrates that large-scale merger-induced sloshing motions are responsible for driving and transporting kinetic energy in a highly coherent fashion, coupling the dynamics of the cluster core and outskirts.
        While some modest variations are present -- for example, the cube outside the merger plane (red) exhibits somewhat lower power, while the region with a visible KHI (orange) displays a heightened normalization -- the normalization between the most and least energetic cubes varies by a factor of $\sim$30, and their spectral slopes are nearly identical. These differences are expected given the local dynamical environment: regions away from the central sloshing plane typically experience less intense gas motions, while sites near visible instabilities such as KHI can show locally enhanced velocities. Nevertheless, the near-uniform spectral shapes and normalization across these diverse environments underscore the coherent, global impact of sloshing motions throughout the cluster core and outskirts.
        
        We find a power-law scaling of $P_k \propto k^{-6}$, corresponding to an energy spectrum of $E(k) \propto P_k k^2 \propto k^{-4}$. This slope is steeper than that of the classical Kolmogorov turbulent power spectrum ($P_k \propto k^{-11/3}$) for an incompressible isothermal gas. The steep slope indicates that the kinetic energy is dominated by large-scale modes, with little power on small scales. 
        This reflects the character of the motions generated in our setup: the coherent sloshing flows and associated shear dominate over smaller-scale velocity fluctuations, consistent with a regime where a turbulent cascade has not fully developed. The steepening of the power spectrum could also be partially attributed to gravitational stratification.

        In contrast, when AGN jet feedback is active, the resulting power spectrum undergoes a significant transformation, though this effect remains highly localized. In the three outlying regions -- the KHI, CF (cold front), and off-plane cubes -- the amplitude and slope of the velocity spectra remain almost unchanged from the sloshing-only simulation, underlining the spatially restricted influence of the AGN. However, in the central cube, encompassing the AGN and its immediate surroundings, the velocity power spectrum slope is significantly flatter, closer to the Kolmogorov expectation. This shallower spectrum indicates an enhancement of kinetic energy at smaller spatial scales in the central region than in the case without AGN feedback, which likely arises from a combination of bulk flows generated by AGN activity (such as cavity inflation) and the  development of turbulence, leading to a richer and more complex velocity field than seen in sloshing alone. Determining the precise decomposition of the velocity field into bulk-flow and turbulent components requires filtering of the velocity field \citep[][]{Dolag2005,Vazza2006,Vazza2012,Schmidt2016,Valdarnini2019,vortexp2024}, which is made non-trivial by the presence of sharp velocity gradients at cold fronts and bubble edges---we leave a detailed analysis of this question for future study. 

        In Figure~\ref{fig:velocity_slices}, we show slices of the velocity field revealing the interplay between turbulent motions inside the AGN-inflated bubbles (visible as enhanced velocities in the $z$ direction) and coherent bulk flows along the $y$ direction, coinciding with the jet axis.

        Together, these findings show that sloshing alone produces predominantly large-scale, coherent motions, while AGN jets generate localized disturbances that inject power across a broader spectrum of scales and strongly enhance velocity fluctuations in the central ICM. This further reinforces the conclusion from the velocity dispersion analysis: AGN feedback is required to account for the small-scale motions observed near the center of the Perseus cluster.

    \subsection{Line-of-sight Velocity Structure}\label{sec:losv}

        \begin{figure*}[ht!]
            \centering
            \includegraphics[width=1\linewidth]{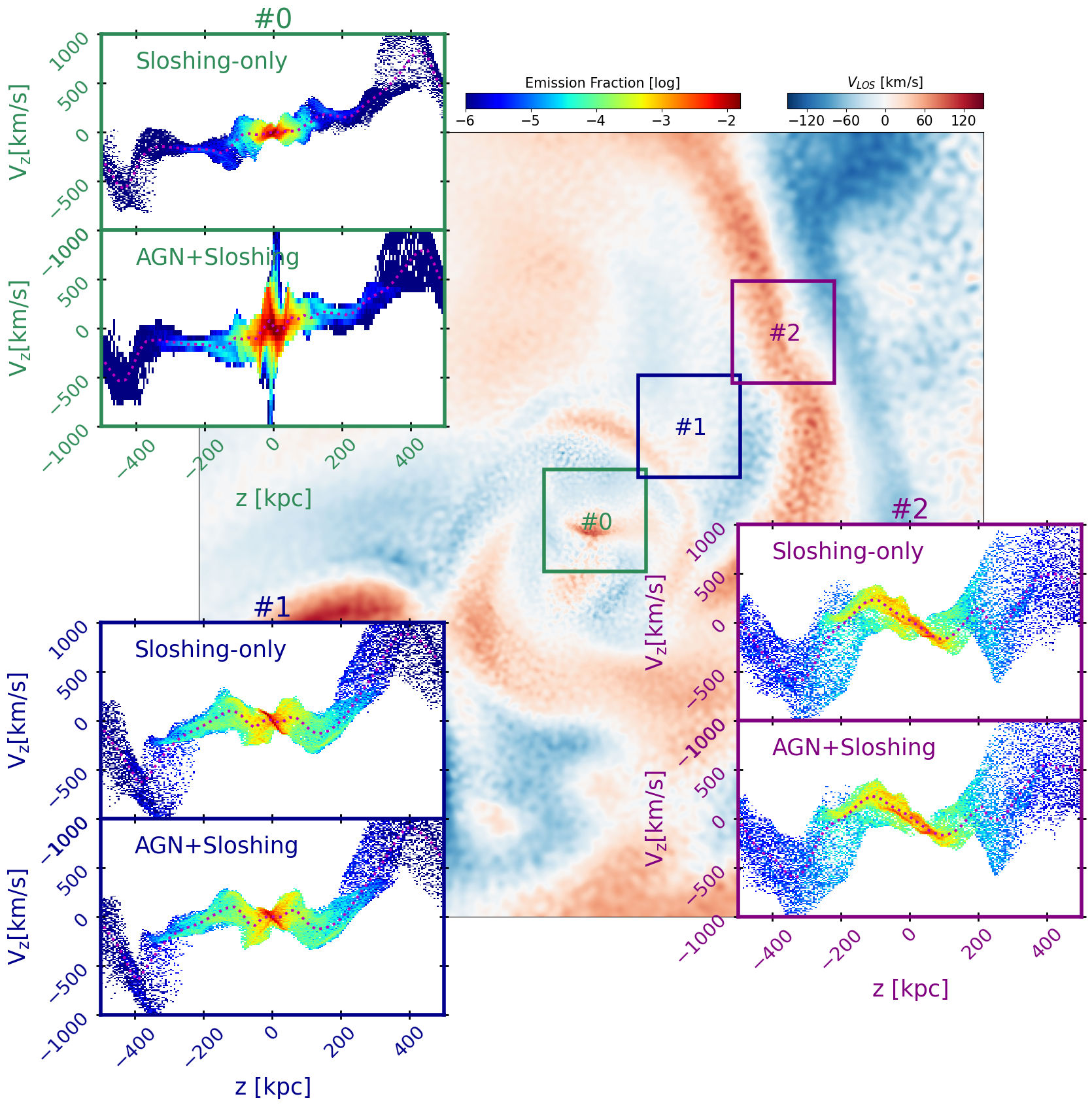}
            \caption{
            \textsl{Velocity structure} of three pointings. The background shows a map of the projected bulk velocity field along the line of sight, with three $60\times60$ kpc$^2$ XRISM-like pointings overlaid. The green pointing is centered on the potential minimum, representing the core, while the blue and purple pointings are positioned along the diagonal arm, following the observational approach of \citep{xrism_perseus_2025}. The inset panels display two-dimensional histograms of the X-ray emissivity-weighted line-of-sight velocity ($v_z$) as a function of projected distance along the sightline ($z$), in the $6–8$~keV band, extracted from prisms 1~Mpc deep. The magenta line is the weighted average profile of the LOS velocity along the line of sight. Panel frames correspond to each pointing's color; the top row presents the velocity structure for the sloshing-only case, while the bottom row shows the combined sloshing and AGN jet scenario.}
            \label{fig:velstruct_differentpointings}
        \end{figure*}

        A central question is how gas motions driven by AGN feedback and merger-induced sloshing manifest in observable X-ray spectral features, in particular, the detailed shapes of emission lines as detected by instruments such as XRISM. In this section, we investigate how sloshing and AGN feedback each imprint signatures on both the velocity structure of the intracluster medium (ICM) and the resulting X-ray line profiles.

        By visualizing the spatial distribution and characteristics of gas velocities, we can identify the distinct patterns produced by sloshing and AGN activity. We then explore how these underlying velocity structures translate into observable differences in emission line profiles, highlighting the unique fingerprints that each driving mechanism leaves on the X-ray spectrum of the cluster.

         To characterize the distribution of gas velocities along the line of sight within the typical XRISM Resolve field of view, we analyze the central region of the Perseus Cluster, corresponding to a projected area of $60 \times 60$~kpc at Perseus' distance. We extract data from a rectangular prism centered on the gravitational potential minimum of our simulated cluster (green panel in Figure~\ref{fig:velstruct_differentpointings}), matching this footprint and extending 1~Mpc along the line of sight ($z$-axis). We also do so for two other regions along an arm extending to the NW, similar to the arm of pointings analyzed in \citet{xrism_perseus_2025}. Within these volumes, we construct a two-dimensional histogram of line-of-sight velocity $v_z$ as a function of $z$, with each value weighted by the local X-ray emissivity in the $6-8$ keV band. This procedure allows us to assess how the velocity structure along the line of sight influences the resulting X-ray emission features observed by XRISM.
         The background of Figure~\ref{fig:velstruct_differentpointings} displays the projected bulk velocity along the sight line, highlighting the three regions under analysis. The inset panels show the line-of-sight velocity structure within these areas, presented as X-ray emissivity–weighted histograms. A similar technique was utilized in \citet{zuhone_astroh_2016} to make predictions for future measurements of gas sloshing velocities by \textit{Hitomi}.

        We first focus on the central pointing and the corresponding velocity structures in the top left panels, framed in green.
        The top panel of the two shows the sloshing-only case, based on the same simulation snapshot discussed previously. We focus here on the line of sight perpendicular to the plane of the merger. The most distinctive feature is the ``wing-like'' shape of the velocity structure, a hallmark of sloshing-induced motions in the ICM first described by \citet{zuhone_astroh_2016}. These wings arise because the cold front forms expanding shell-like structures in three dimensions, extending both in and out of the plane of the sky as it moves away from the cluster center. Specifically, gas in the cold front located behind the cluster core (with respect to the observer) moves away from the observer, while gas in front moves toward the observer. This geometry produces a pair of broadened, quasi-symmetric wings in the velocity–distance distribution.

        The color scale in the figure represents X-ray emissivity, highlighting that the gas within these wings is of lower density and therefore less luminous in X-rays; its overall contribution to the observed emission is minor. By contrast, the densest and most X-ray–bright gas, which dominates the emission line signal, resides near the cluster center, where the LOS velocities are generally quite small. Thus, while sloshing leads to significant bulk flows and shear throughout the ICM, these do not generate substantial LOS velocity dispersions in the regions that contribute the majority of the X-ray emission. Consequently, in the sloshing-only case, the largest line-of-sight velocities occur in X-ray faint regions, while the X-ray–bright gas remains near zero velocity (in the rest frame of the dark matter halo); thus, sloshing fails to produce significant LOS velocity dispersion in the cluster center.

        In contrast, 
        the bottom green panel of Figure~\ref{fig:velstruct_differentpointings} shows the same diagnostic for the simulation including AGN feedback. The larger scale ``wing-like'' features in the velocity structure, arising from sloshing, remain largely unchanged by the central AGN jets. However, the behavior in the central region is markedly different: gas within $z \in [-30, 30]$~kpc exhibits a much broader spread of LOS velocities, with extremes reaching beyond 1000~km~s$^{-1}$. Notably, the X-ray–brightest material reaches line-of-sight speeds up to 500~km~s$^{-1}$, a clear contrast to the much narrower range seen in the sloshing-only case.
        This broadening reflects the presence of substantial velocity components generated by AGN-driven activity, such as cavity inflation and the associated displacement of surrounding gas, which -- even though it injects kinetic energy predominantly along the jet axis -- ultimately drives motions in multiple directions within the core. 

        It is noteworthy that in this simulation, AGN jets are injected along the $y$-axis, lying in the plane of the sky and perpendicular to the line of sight under consideration. Despite this orientation, we still observe significant LOS velocity components in the X-ray bright regions. This highlights the ability of AGN feedback to induce complex, multi-directional gas motions within the cluster core, and demonstrates that such activity can strongly shape the velocity structure even when the direct jet axis is not aligned with our line of sight. We discuss the scenario in which the line of sight is aligned with the jet axis in Section \ref{sec:line_profiles}.

        The blue and purple inset panels in Figure~\ref{fig:velstruct_differentpointings} display the velocity structure in two additional pointings along the NW diagonal, moving progressively farther from the cluster center. In each of these locations, the characteristic ``wing-like'' patterns produced by sloshing are clearly visible in the two-dimensional distributions of line-of-sight velocity. We observe that moving away from the center, the differences between the sloshing-only and AGN+sloshing simulations become increasingly subtle. While AGN-driven turbulence has a pronounced impact on the velocity structure in the cluster core, its influence rapidly diminishes with radius. The velocity structure at the most external pointing (\#2) is virtually indistinguishable between the two cases, hence dominated by the large-scale coherent motions induced by sloshing, and the imprint of AGN feedback becomes negligible. This clearly shows that while AGN activity strongly influences the velocity field in the center, its signature quickly diminishes outside the core, e.g., $r \sim 50-100$~kpc. A similar conclusion regarding the region affected by AGN feedback was reached in tailored jet simulations and density fluctuation studies by \citet{heinrich_2021}.

        \begin{figure*}[ht!]   
        \centering
        \includegraphics[width=1\textwidth]{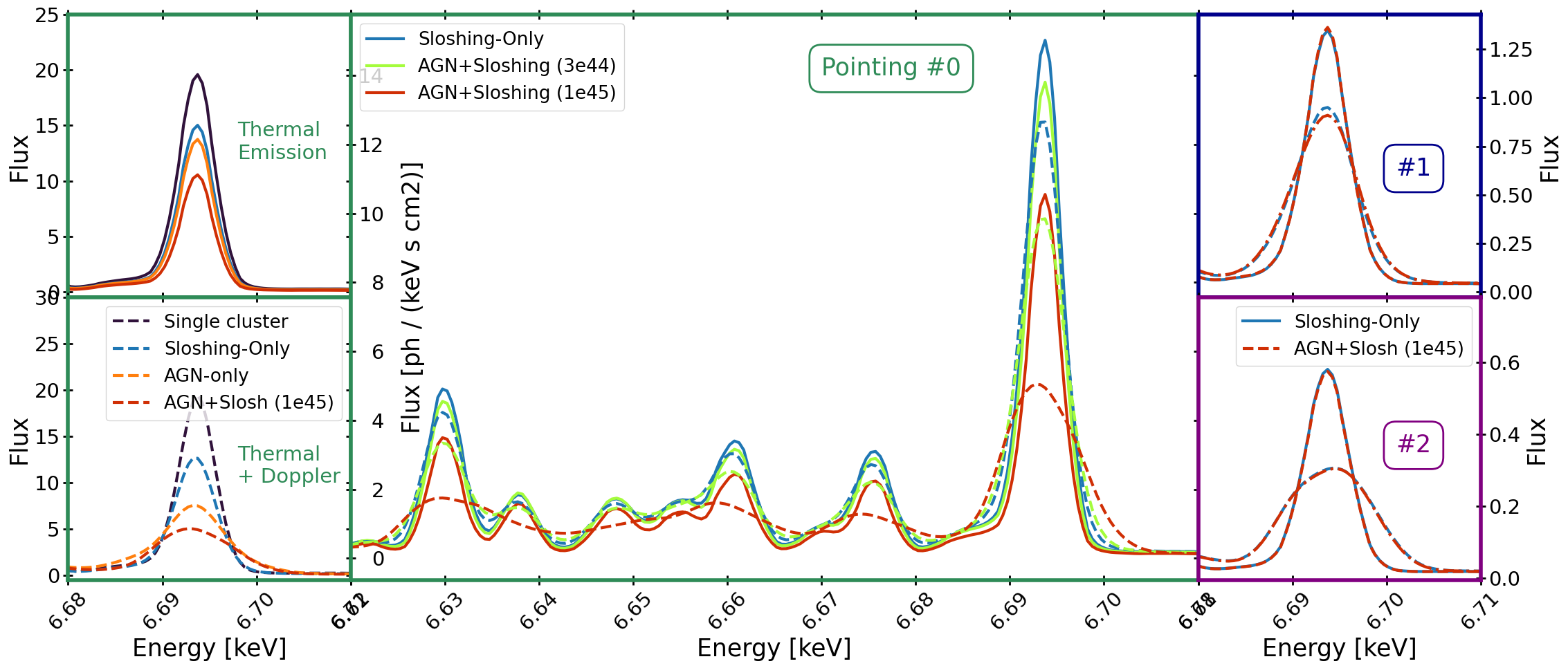}
        \caption{\textsl{Synthetic line profiles} for the 6.7 keV Fe XXV line computed within $60\times60\times1000$ kpc$^3$ volumes centered on the three pointings shown in Figure~\ref{fig:velstruct_differentpointings}. The two panels on the left and the large central panel focus on the central pointing. Solid curves show emission profiles broadened by thermal motions only, while dashed curves include both thermal and line-of-sight velocity broadening. The left panels compare the line broadening produced in a relaxed, single cluster (no sloshing), a sloshing-only case, an AGN-only case with $P_{\mathrm{jet}} = 1 \times 10^{45}$~erg~s$^{-1}$, and a simulation with both AGN+sloshing (same jet power). The central panel presents the Fe line complex for the sloshing-only case, as well as for two different AGN jet powers ($3 \times 10^{44}$~erg~s$^{-1}$ and $1 \times 10^{44}$~erg~s$^{-1}$), illustrating the effect of increasingly powerful jet activity. The two rightmost panels show the Fe line profiles corresponding to the additional off-center pointings along the diagonal.
        All the fluxes are in photon~keV$^{-1}$~s$^{-1}$~cm$^{-2}$.
        \label{fig:line_profile}}
        \end{figure*}

        \begin{figure*}[ht!]   
            \centering
            \includegraphics[width=0.85\textwidth]{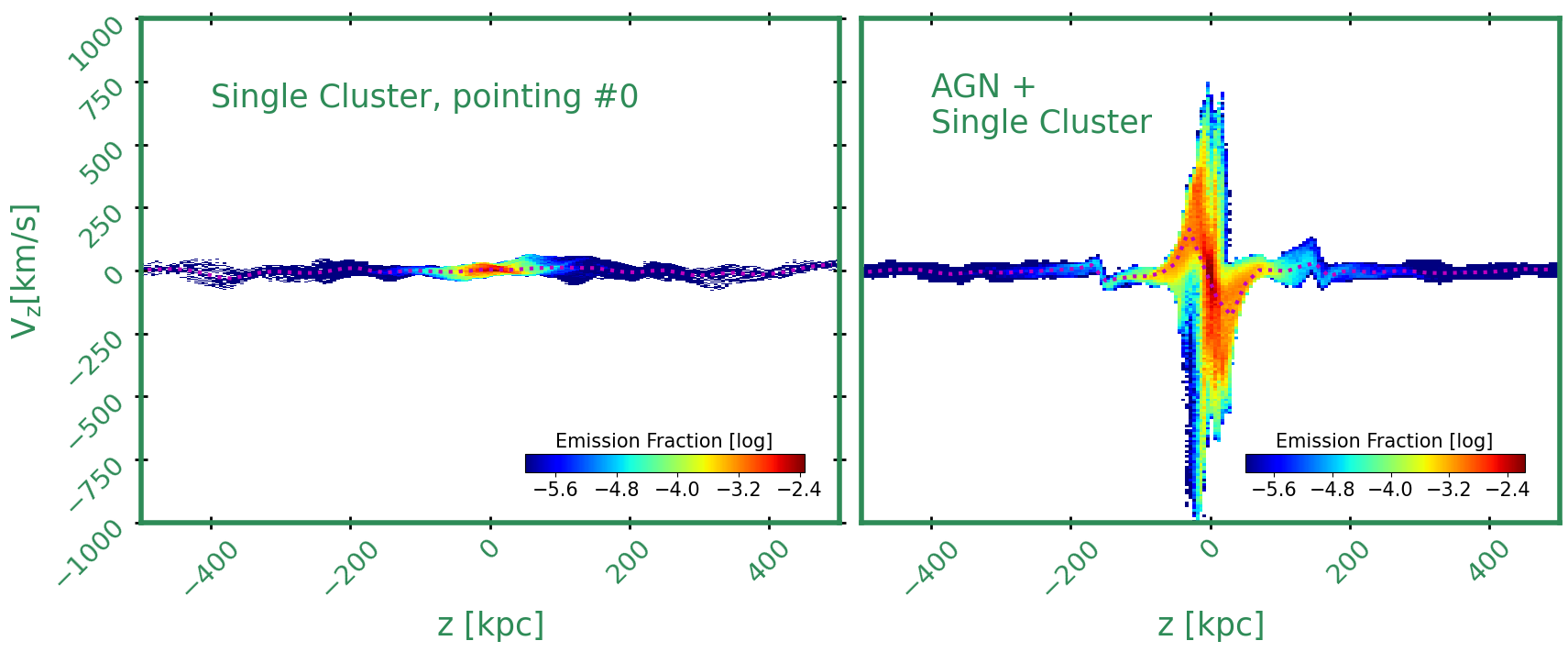}
            \caption{Velocity structure for simulations of a single cluster without sloshing. Emissivity-weighted 2D histograms of the velocity along the line of sight ($v_z$) as a function of position along the LOS ($z$) within a 60$\times$60~kpc$^2$ region centered on the cluster core and extending 1000~kpc along the line of sight. The \textit{left panel} shows a relaxed cluster with no AGN activity, while the \textit{right panel} includes AGN jet feedback with a $1 \times 10^{45}$~erg~s$^{-1}$. In both cases, the lack of merger-induced sloshing results in the absence of the characteristic wing-like patterns seen in previous figures. 
            }
            \label{fig:singlecluster_velstructure}
        \end{figure*}

        \subsection{Line Profiles}\label{sec:line_profiles}
        To understand how the velocity distribution within the cluster core affects observable spectral features, we now examine the Fe~XXV emission line complex at a rest-frame energy of $\sim$6.7~keV. Using the same volume as the central pointing of Figure~\ref{fig:velstruct_differentpointings} 
        $-$ a rectangular prism of $60\times60\times1000$ kpc$^3$ centered on the cluster potential minimum $-$ we compute synthetic line profiles by summing the spectral contribution of each gas cell along the line of sight. The emissivity of each cell is used to weight its contribution, and the thermal broadening of the line is calculated based on the temperature of each cell. These purely thermally broadened spectra are shown as blue and green lines in Figure~\ref{fig:line_profile}.
        To capture the full dynamical imprint of the ICM motions, we then incorporate the Doppler shift introduced by each gas cell's velocity along the line of sight. The resulting profiles, which account for both thermal and kinematic effects on the spectrum, are plotted in orange and red. In all simulations, the line is broadened beyond the thermal profile, and to some degree shifted, depending on the net LOS bulk motion.

        To isolate the individual contributions of sloshing and AGN feedback to the observed velocity structures and line profiles, we run a pair of simulations in which no merger occurs and the cluster remains at rest. One simulation includes AGN jet activity, while the other features a relaxed, quiescent core. These setups allow us to disentangle the effects of jet-driven motions from those induced by large-scale sloshing. The results are shown in Figure~\ref{fig:singlecluster_velstructure}, which presents 2D histograms of the line-of-sight velocity as a function of position along the LOS.

        In the absence of both sloshing and AGN activity (left panel in Figure~\ref{fig:singlecluster_velstructure}), the gas velocities remain extremely low throughout the volume, with 90\% of the speeds being lower than $\sim25$~km~s$^{-1}$. The emissivity-weighted velocity structure lacks the characteristic ``wing'' patterns seen in sloshing simulations, and this quiescent state translates directly to the line profile were the thermal-only and the one including the Doppler broadening are virtually indistinguishable (see dark purple lines in the left panels of Figure~\ref{fig:line_profile}).

        By contrast, the case with AGN feedback but no sloshing (right panel of Figure~\ref{fig:singlecluster_velstructure}) exhibits high LOS velocities localized to the cluster core, where the jets are injecting energy. Importantly, the absence of extended ``wings'' in the 2D histogram confirms that these features cannot be explained by center perturbations like AGN feedback that eventually expand at larger radii.        
        While the outer regions remain kinematically quiet, the central jet-inflated cavities drive gas motions that lead to noticeable velocity structures near the center, which are then reflected into the broadening of the line profile (see orange lines in the left panels of Figure~\ref{fig:line_profile}). 
        This reinforces the idea that AGN feedback can generate substantial velocity dispersion in the X-ray–bright regions even in the absence of merger-driven sloshing.

        Having established the baseline line profiles and velocity structures resulting from each process in isolation, we now turn to the full cluster simulations where sloshing, AGN feedback, or both mechanisms operate in concert. We first focus on the line profiles (see Figure~\ref{fig:line_profile}) extracted from the central pointing (the green region in Figure~\ref{fig:velstruct_differentpointings}), which captures the core of the cluster. Considering only thermal broadening (top left panel), the single, relaxed cluster case produces the highest total line emission, followed by the sloshing-only and AGN-only configurations, both of which displace some bright, dense gas from the cluster center. The combined sloshing plus AGN scenario yields the lowest central emission, as expected due to the combined redistribution of gas by both processes. When the effect of line-of-sight velocities (Doppler broadening, bottom left panel) is included, we see that sloshing alone already widens the profile, consistent with findings from \citet{zuhone_astroh_2016}; however, because the X-ray–bright core produced by sloshing contains only modest LOS velocities, this broadening remains relatively limited. In contrast, including AGN feedback causes a more pronounced increase in line width -- particularly in the AGN+sloshing case (shown by the dashed lines) -- as the injection of AGN-driven bulk motions and turbulence populates the core with high-velocity gas.

        As the pointings move away from the cluster center (blue and purple pointings in Figure~\ref{fig:velstruct_differentpointings}), the total X-ray emission naturally decreases (right panels in Figure~\ref{fig:line_profile}), reflecting the declining gas density in the outskirts and consequently the X-ray flux. Additionally, in these off-center regions, the synthetic line profiles produced by the sloshing-only and AGN+sloshing simulations become nearly indistinguishable. This similarity indicates that, outside the core, the kinematic imprint of AGN-driven motions is minimal and the velocity structure is dominated by merger-induced sloshing.

        To gain further insight into the impact of gas motions on X-ray line profiles, we now consider lines of sight within the plane of the merger -- that is, the $x$ and $y$ directions, rather than the $z$ direction, which is perpendicular to both the merger plane and the direction of the AGN jets as previously discussed. This change of perspective provides a more complete view of the kinematic signatures introduced by both AGN activity and sloshing within the cluster core.

        \begin{figure*}[ht!]   
        \centering
        \includegraphics[width=0.63\textwidth]{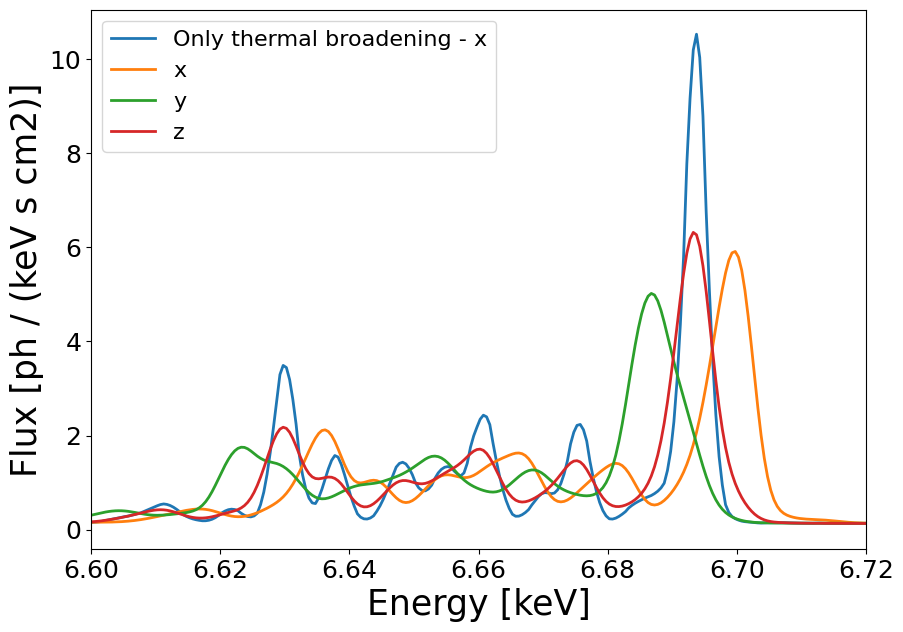}
        \includegraphics[width=0.36\textwidth]{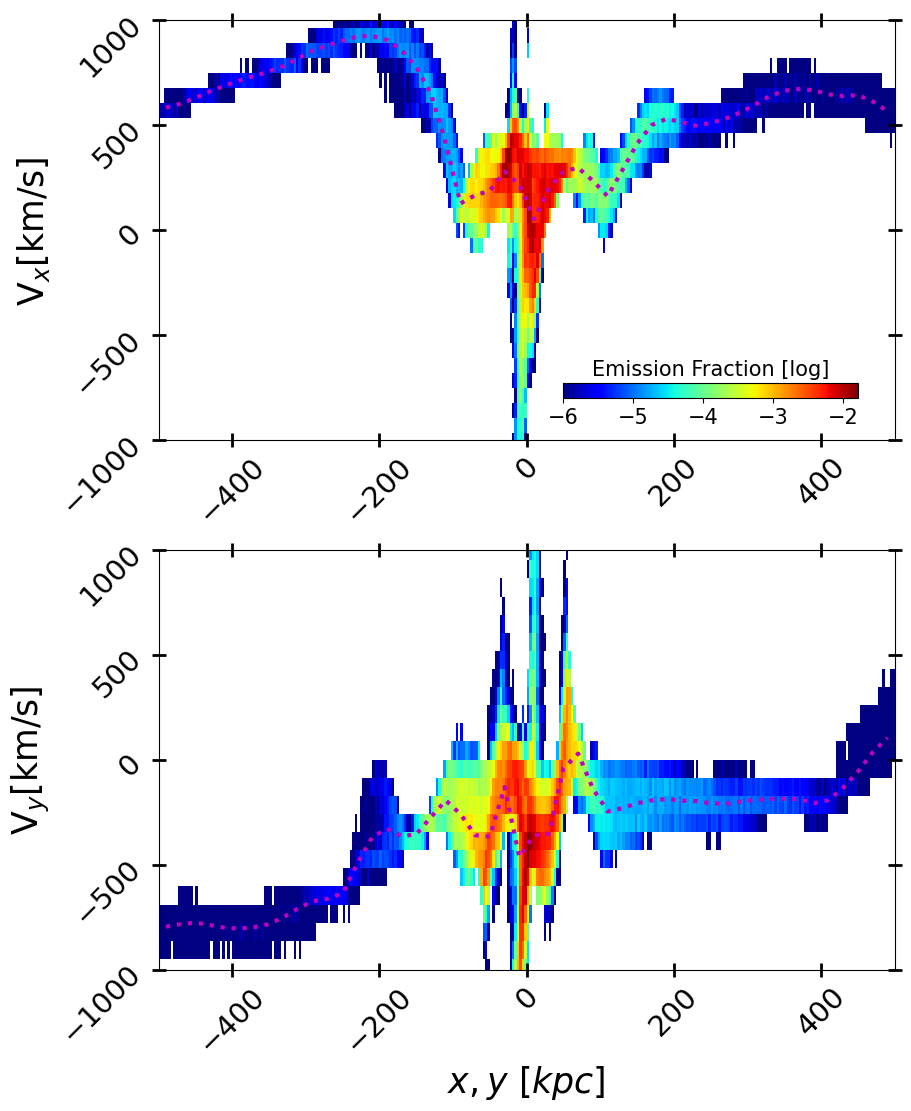}\\
        \includegraphics[width=0.49\textwidth]{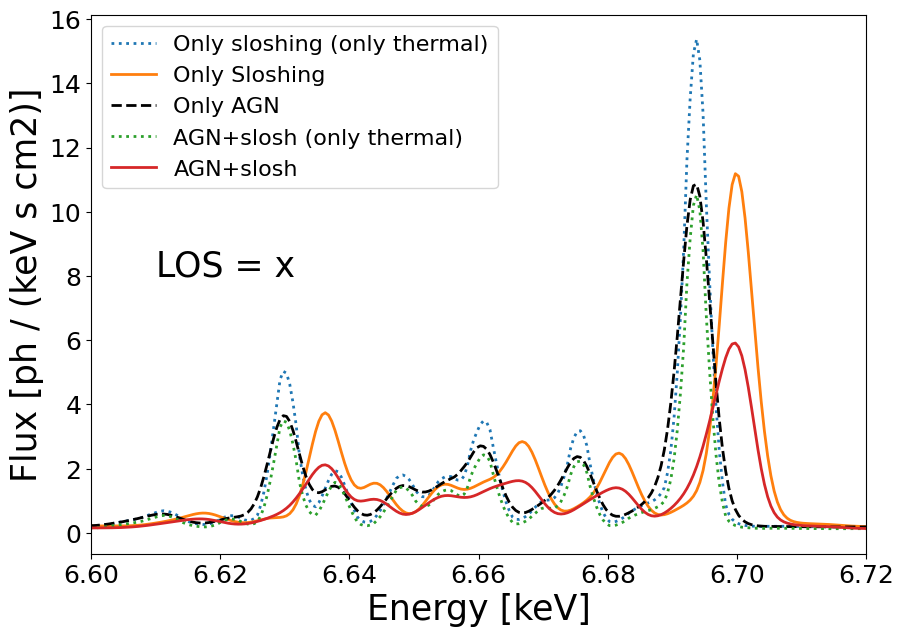}
        \includegraphics[width=0.49\textwidth]{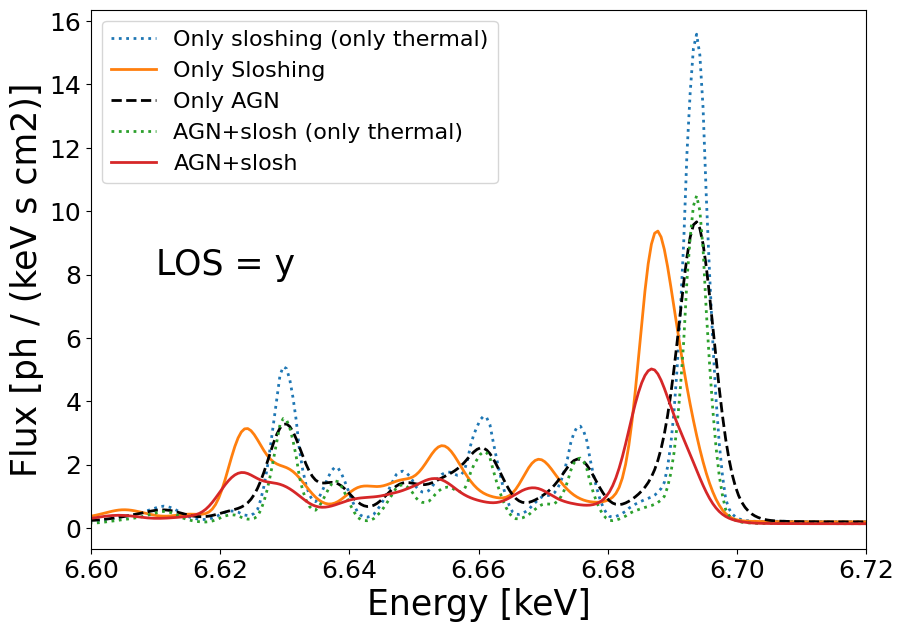}
        \caption{\textit{Top left:} Line profiles from the AGN+sloshing simulation, showing thermal broadening only (blue curve) and the combined effects of thermal broadening and Doppler shifts for observers located along the $x$, $y$, and $z$ axes. 
        \textit{Top right:} X-ray weighted velocity structures projected along the $x$-axis (upper right) and $y$-axis (lower right), illustrating the gas velocity structure for observers in those directions. 
        \textit{Bottom panels:} Comparison of emission line profiles for the three physical scenarios -- sloshing only, AGN only, and AGN+sloshing, shown for observers positioned along the $x$ and $y$ directions.}
        \label{fig:line_profile_xy}
        \end{figure*}

        As shown in Figure~\ref{fig:line_profile_xy}, an observer viewing the system from the $x$ or $y$ directions would notice not only a broadening of the emission lines due to the influence of the jets, but also a clear shift in the line centroid due to the sloshing. In the top left panel, we compare the line broadening and shift for the central pointing in the simulations with AGN feedback on top of the sloshing motions.
        The broadest line profiles are found for observers looking along the $y$-axis (green line), which is aligned with the jet direction. This is expected, as the jets induce strong motions along this axis, thereby widening the line profile through Doppler effects. The symmetry of the bi-directional jet prevents it from contributing to significant line shifts by itself. The observed line shifts depend strongly on the viewing angle. The line observed along the $x$-axis (orange line) is blue-shifted, while the line observed along the $y$-axis (green line) is moving away from the observer and so red-shifted. These line shifts are largely produced by sloshing motions, which are asymmetric and thus produce a net bulk velocity in projection.

        The bottom panels of the figure allow for a direct comparison between the sloshing-only, AGN-only, and combined scenarios (AGN+sloshing). When only AGN feedback is present and sloshing is absent, the bubbles inflated by the jets remain almost perfectly aligned along the jet axis. In this case, the emission lines are noticeably broadened by the expansion of the bubbles, but exhibit minimal net Doppler shift, the black curves remain centered on the rest energy for all lines of sight. As already noted, this indicates that, on their own, jet-driven bubbles do not impart significant bulk motion to the X-ray–emitting gas in any particular direction, but rather broaden the line profile due to their expansion velocities. In contrast, when both AGN activity and sloshing are present, the sloshing motions not only produce observed bulk velocities on their own, but also bend the jets and displace the bubbles in the plane of the merger (e.g., \citealt{paola_2024}), introducing asymmetry as well in the AGN-driven motions.  These combined effects on the shifting and broadening of the emission lines are most pronounced for observers looking within the merger plane, where the interplay between jet-driven motions and sloshing-induced displacements generates complex, direction-dependent line profiles. 
        
        Additionally, when viewed along directions within the merger plane, the line profiles can show deviations from a purely Gaussian shape, reflecting the projection of multiple velocity structures projected along the same sight line. This effect is present in the case of pure sloshing motions \citep[orange solid curves, see also][]{zuhone_astroh_2016}, but is even more pronounced when both AGN and sloshing-driven motions are present.

\section{Conclusion}\label{sec:conclusions}

        Understanding the origins of gas motions in the intracluster medium (ICM) is a central challenge in the study of galaxy clusters, with broad implications for cluster evolution, feedback processes, and the microphysics of the hot plasma. Recent advances in high-resolution X-ray spectroscopy, especially with missions like XRISM, have opened a new window into measuring cluster velocity structure directly, providing crucial observational constraints on the sources and nature of turbulence and bulk flows. However, interpreting these data requires disentangling the relative contributions of various physical processes -- most notably, feedback from active galactic nuclei (AGN) and merger-driven sloshing -- that can inject energy and drive complex velocity fields across a range of spatial scales.
        
        In this work, we have leveraged a suite of idealized, controlled simulations ran with AREPO and tailored to the Perseus cluster to disentangle the effects of AGN feedback and merger-driven sloshing on the kinematic properties of the ICM. By isolating these two main physical processes, we have directly assessed their separate and combined impacts on the observed velocity profiles reported by \citet{xrism_perseus_2025} -- 
        most notably, the increased velocity dispersion in the central $\sim$60 kpc and a similar level at larger radii. 
 
        Our results demonstrate that neither AGN activity nor sloshing alone can reproduce the full complexity of the observed velocity dispersion profile. Sloshing motions -- regardless of subcluster mass or trajectory -- primarily drive large-scale, coherent flows and account for the elevated velocity dispersions at larger radii, but they systematically underestimate the observed central dispersion within the cluster core. In contrast, AGN feedback, especially at higher jet powers, efficiently injects bulk motions and turbulence into the core and raises the central velocity dispersion, but fails to sustain enhanced dispersions in the outskirts. Only by combining both mechanisms do we recover a velocity dispersion profile that qualitatively matches the structure observed in Perseus, with elevated velocity dispersion in the cluster center and the outskirts, with AGN jets dominating the inner $\sim$60 kpc and merger-driven sloshing prevailing beyond, consistent with the conclusions of \citet{xrism_perseus_2025}.
        
        The analysis of the velocity power spectra reveals further evidence for the distinct roles of AGN and sloshing. In simulations with both processes, the inner regions exhibit enhanced power on small and intermediate spatial scales due to jet-driven gas motions, while the outer regions are marked by steep spectra dominated by large-scale coherent motions from sloshing. This spatial variation means that the kinematics of the ICM cannot be described by a single, universal velocity power spectrum, reinforcing the necessity for multi-component models when interpreting spatially resolved velocity data.
        
        Examining the velocity field along the line of sight, we find that AGN feedback produces complex, multi-directional gas motions and broadens the LOS velocity distribution in the cluster core, leading to strongly broadened and sometimes multi-component X-ray emission lines. Conversely, sloshing, while generating substantial internal flows and cold fronts, does not yield significant LOS velocities where most X-ray emission originates, resulting in narrower line profiles. These findings underscore the importance of considering both physical drivers when interpreting high-resolution X-ray spectra from missions such as XRISM. In sum, we support the evidence of at least two drivers with tailored AGN+sloshing feedback simulations -- small-scale AGN feedback and large-scale merger-induced sloshing -- operating in concert to shape the velocity structure of the ICM in clusters like Perseus, as argued in \citet{xrism_perseus_2025} using simple theoretical arguments and driven-turbulence simulations. This has immediate consequences for the modeling and interpretation of X-ray emission line profiles as probes of cluster physics.

        While we have focused here on a specific realization of AGN jet feedback, our main finding is more general: reproducing the observed velocity dispersion profile in Perseus requires an additional central source of kinetic energy beyond merger-induced sloshing. Our results do not imply that the implemented feedback model is unique or fully realistic, but rather highlight the necessity for some AGN-driven process to account for the data. Our results are consistent with the findings of previous studies, which similarly explored the impact of AGN feedback on ICM velocity structure. For example, \citet{Bourne_2017} and \citet{Ehlert_2021} concluded that AGN jets can effectively stir turbulence in the vicinity of the jets and bubbles, though they cannot drive substantial gas motions at larger radii within the cluster -- a result that our analysis also clearly demonstrates. Similarly, \citet{Reynolds2015} and \citet{Yang2016} found that AGN are very inefficient at driving turbulence in clusters outside their immediate environs. 
        
        We also emphasize that both episodic AGN activity and tailored modeling of Perseus’s large-scale sloshing structures are essential for reproducing the prominent AGN contribution to the radial profile of velocity dispersion in the Perseus core region, as seen in the XRISM observations. These aspects represent the major differences from some recent similar studies \citep[e.g.][]{Ehlert_2021,Li2025} and lead to different conclusions about the role of AGN feedback in driving turbulence. \citep[][]{xrism_perseus_2025} indicates that only $\lesssim20\%$ of the observed total kinetic energy arises from large-scale motions, a result supported by the comparison between our sloshing-only and sloshing+AGN simulations. This, in turn, constrains the level of velocity dispersion that can be attributed to pre-existing motions.
        
        Looking ahead, further insight into cluster velocity structure is anticipated from several forthcoming studies based on cosmological simulations. Truong et al. (in preparation) are analyzing a sample of Perseus-like clusters from the TNG-Cluster suite to explore the origin of velocity dispersion profiles in a fully cosmological context and their relation to recent XRISM findings. Likewise, Weinberger et al. (in preparation) are investigating the similarities and differences between self-regulated isolated cluster simulations and cosmological zoom-in runs employing the same jet feedback models. The results of these studies will provide a valuable complement to our idealized experiments, offering the opportunity to place our conclusions in the context of more complex and self-consistent cluster environments. Together, these efforts will help to further clarify the role of AGN feedback and merger-driven motions in shaping the velocity structure of the ICM.

\software{astropy \citep{Astropy2013,Astropy2018,Astropy2022},  
yt \citep{Turk2011}, matplotlib \citep{Hunter2007}, numpy \citep{Harris2020}, scipy \citep{Virtanen2020}.}


\begin{acknowledgments}
{Support for EB was provided by the NASA XRISM Guest Scientist contract 80NSSC23K0751. Support for JAZ was provided by the {\it Chandra} X-ray Observatory Center, which is operated by the Smithsonian Astrophysical Observatory for and on behalf of NASA under contract NAS8-03060. 
RW acknowledges funding of a Leibniz Junior Research Group (project number J131/2022). 
CP acknowledges support from the European Research Council via the ERC Advanced Grant ``PICOGAL'' (project ID 101019746) and by the Deutsche Forschungsgemeinschaft (German Research Foundation) for the Research Unit FOR5195 on Relativistic Jets in Active Galaxies (443220636). 
MR acknowledges support from the National Science Foundation Collaborative Research Grant NSF AST-2009227 and NASA ATP grant 80NSSC23K0014.
JHL acknowledges the support from the Canadian Space Agency through the Financial Support for XRISM Guest Scientist (XGS) Opportunity. 
MLGM acknowledges financial support from NSERC via the Discovery grant program and the Canada Research Chair program. The material is based upon work supported by NASA under award number 80GSFC24M0006.
CZ acknowledges the support of the Czech Science Foundation (GACR) Junior Star grant no. GM24-10599M. IZ acknowledges support from the NASA grant 80NSSC18K1684 and partial support from the Alfred P. Sloan Foundation through the Sloan Research Fellowship.
We acknowledge the use of OpenAI's ChatGPT for assistance in editing this manuscript for clarity and conciseness and for making minor edits to codes.}
\end{acknowledgments}




\bibliography{biblio}{}
\bibliographystyle{aasjournal}



\end{document}